\documentclass[11pt]{article}
\pdfoutput=1
\usepackage{jheparxiv}

\usepackage[dvipsnames]{xcolor}

\usepackage{bm}
\usepackage{amsmath}
\usepackage{amssymb}
\usepackage{graphicx}
\usepackage{slashed}
\usepackage{amsfonts}
\usepackage{comment}
\usepackage{cancel} 
\usepackage{hyperref}
\hypersetup{
colorlinks=true,         
linkcolor=blue,          
citecolor=red,        
urlcolor=blue            
}

\newcommand{\be}{\begin{equation}}
\newcommand{\ee}{\end{equation}}
\newcommand{\bi}{\begin{itemize}}
\newcommand{\ei}{\end{itemize}}
\newcommand{\bea}{\begin{eqnarray}}
\newcommand{\eea}{\end{eqnarray}}
\newcommand{\ud}{\mathrm{d}}
\newcommand{\LCm}{{\scriptscriptstyle -}} 
\newcommand{\LCp}{{\scriptscriptstyle +}}
\newcommand{\LCpm}{{\scriptscriptstyle \pm}}
\newcommand{\LCmp}{{\scriptscriptstyle \mp}}
\newcommand{\LCperp}{{\scriptscriptstyle \perp}}

\newcommand{\deltahat}{\hat{\delta}}
\newcommand{\Epsilon}{\mathcal{E}}

\begin{document}

\title{Scattering amplitudes and electromagnetic horizons}

\author{Anton Ilderton}
\emailAdd{anton.ilderton@ed.ac.uk}

\author{and William Lindved}
\emailAdd{william.lindved@ed.ac.uk}

\affiliation{Higgs Centre, University of Edinburgh, EH9 3FD, Scotland, UK}

\abstract{
We consider the scattering of charged particles on particular electromagnetic fields which have properties analogous to gravitational horizons. Classically, particles become causally excluded from regions of spacetime beyond a null surface which we identify as the `electromagnetic horizon'. In the quantum theory there is pair production at the horizon via the Schwinger effect, but only one particle from the pair escapes the field. Furthermore, unitarity appears to be violated when crossing the horizon, and there is no well-defined S-matrix. Despite this, we show how to use the perturbiner method to construct `amplitudes' which contain all the dynamical information required to construct observables related to pair creation, and to radiation from particles scattering on the background.
}

\maketitle

\section{Introduction}
Modern QFT methods have the potential to greatly simplify the calculation of observables related to classical gravitational waves generated by e.g.~black hole collisions, for reviews see~\cite{Travaglini:2022uwo,Bjerrum-Bohr:2022blt,Kosower:2022yvp,Buonanno:2022pgc}. Within the amplitudes approach to gravitational scattering, gravity is treated as an effective theory valid below the Planck scale, and black holes are treated as point-like masses, thus without internal structure or horizons; for large-distance scattering, implying weak coupling, this is clearly sufficient. The extraction of observables from amplitudes is then facilitated by techniques such as double copy~\cite{Bern:2019prr,Bern:2022wqg,Adamo:2022dcm}, generalised unitarity~\cite{Bern:2011qt}, and a systematic approach to taking the classical limit~\cite{Kosower:2018adc,Cristofoli:2021vyo}.  
A prominent and challenging line of research is to push perturbative post-Minkowskian (`PM') scattering calculations~\cite{Neill:2013wsa,Damour:2016gwp,Cheung:2018wkq} to higher order in $G_N$~\cite{Bern:2019nnu,Bern:2021dqo,Bjerrum-Bohr:2022blt}, in order to improve precision for comparison with future experiments~\cite{Punturo:2010zz,LISA:2017pwj,Kalogera:2021bya}.

There are many other open questions and challenges. Can we apply scattering results to systems with bound orbits, i.e.~is there a scattering-to-bound dictionary~\cite{Kalin:2019inp,Kalin:2019rwq,Cho:2021arx,Gonzo:2023goe}?  What happens when quantum effects become larger than classical self-force effects~\cite{Bellazzini:2022wzv}, and we cannot benefit from the simplifications of the classical limit? What techniques can we bring to bear in the presence of strong curvature~\cite{Adamo:2022rmp}, or beyond the regime of weak-field calculations? In spacetimes with horizons, to what extent can we define an $S$-matrix or global observables~\cite{Hawking:1975vcx,Gibbons:1977mu,Witten:1998qj,Bousso:2004tv}, and to what extent can amplitudes methods probe the horizons?

The goal here is to explore analogous questions in the simpler setting of electrodynamics (the usefulness of which as a toy model for PM calculations in gravity has been emphasised in e.g.~\cite{Bern:2021xze,Bern:2023ccb}). We will consider particles scattering on a background electric field, which will play a role analogous to spacetime curvature. The chosen field exhibits an `electromagnetic horizon' which causally excludes regions of spacetime from charged particles. To clarify the type of horizon we consider~\cite{Ashtekar:2004cn}, we note that while there is curvature, it is the acceleration caused by the field which realises the horizon. Indeed, the class of fields we will consider contains that which causes uniform acceleration; we have an electromagnetic system which realises a (non-uniform acceleration) generalisation of the Rindler horizon. Quantum mechanically, we will see that there is particle production at this horizon, and that there is a violation of unitarity when crossing the horizon. We will discuss how this is resolved by analysis of the information content of various surfaces. It is interesting that these issues arise in a system which is, we will see, almost semiclassical-exact, given recent results that motivate the consideration of semiclassical models~\cite{Akers:2019nfi} in regard to the information paradox.

We emphasise that the system we consider is already well-studied in the context of non-perturbative pair production~\cite{Tomaras:2000ag,Tomaras:2001vs,Fried:2001ur}. The methods and perspective presented here are however new, and they will allow us to connect to gravity, to study other scattering processes and present new calculations, and to address some outstanding questions from the literature.

This paper is organised as follows. We begin in Section~\ref{sect:classical} by introducing the system of interest, identifying the electromagnetic horizon, and discussing the extent to which it causally excludes portions of spacetime from charged particles. We also calculate some typical classical observables and self-force effects. In Section~\ref{sect:quantum} we extend these results to the quantum theory by analysing the wavefunctions which describe scalar and spinor fields scattering on our background. We will see that these wavefunctions diverge at the classical horizon, and that as we cross the horizon unitarity is violated. We connect this to particle production in the background via the Schwinger effect and discuss how, in this system, the pair production mechanism realises concretely many properties attributed to Hawking radiation~\cite{Hawking:1975vcx,Gibbons:1977mu}. In Section~\ref{sect:scattering} we turn to the calculation of observables in QFT, using the perturbiner method to recover the known pair production probability, despite there being no $S$-matrix.
 
 We analyse two other situations; the impact of quantum interference effects, and the classical limit of radiative observables from amplitudes on the background. We conclude in Section~\ref{sect:concs}. The Appendices contain results on spin and lightfront zero modes referred to in the text, as well as a discussion of other approaches.

\section{Classical}\label{sect:classical}

We consider a classical scalar electron, mass $m$ and charge $e$, colliding with an electromagnetic wave.
Neglecting, for the moment, radiation (back-reaction on the background) and radiation reaction (self-force) effects, the wave becomes a prescribed and fixed background $F^\text{ext}_{\mu\nu}$, on which the particle moves as a test charge according to the Lorentz force law
\be\label{LorentzForce}
    \ddot{X}_\mu(\tau) = \frac{e}{m}F^\text{ext}_{\mu\nu}(X(\tau)){\dot X}^\nu(\tau) \;,
\ee
in which $X^\mu$ is the particle orbit, parameterised by proper time $\tau$. This is the analogue of geodesic motion. (Radiation and reaction will be reintroduced later.)

It will be convenient to work throughout in lightfront coordinates,
\be
    \ud s^2 = 2 \ud x^\LCp \ud x^\LCm - \ud x^1 \ud x^1 - \ud x^2 \ud x^2 \,,
\ee
in which $x^\LCm$ is `lightfront time' and we write $x^\LCperp=(x^1,x^2)$ for the `transverse' coordinates. In terms of cartesian components we have $x^\LCpm =(x^0 \pm x^3)/\sqrt{2}$. We introduce two null vectors $n$ and ${\bar n}$ such that $n\cdot x = x^\LCm$ and ${\bar n}\cdot x = x^\LCp$. Our chosen background field is
\be\label{TheField}
    A^\text{ext}(x) = -x^\LCp \mathcal{E}(x^\LCm)\ud x^\LCm \;,
\ee
so the only non-zero component of the field strength is $F^\text{ext}_{\LCm\LCp} = \mathcal{E}(x^\LCm)$. For later use we define the classical potential
\be\label{eq:definition-b}
    b(x^{\LCm}) := \int_{- \infty}^{x^{\LCm}}\!\ud y \, e \mathcal{E}(y) \;.
\ee
Physically, (\ref{TheField}) represents an electric field travelling in the positive $x^3$-direction and polarised in the same direction. The field is thus not a vacuum solution, being sourced by a null current $J_\LCm = -\mathcal{E}'(x^\LCm)$, but this is not important for the discussion. We assume that the physical field $\mathcal{E}(x^\LCm)$ is nonsingular and compactly supported in the region $x^\LCm_i < x^\LCm < x^\LCm_f$ (a `sandwich' wave). The field thus splits spacetime into three pieces, with two flat regions, which we label `in' and `out', on either side of the wave, see Fig.~\ref{fig:classical}. All incoming massive particles begin in the in-region, and are eventually struck by the propagating field (at the same lightfront time $x^\LCm=x_i^\LCm$). This field is not `asymptotically flat' in the $x^\LCp$--direction, note. In the limit that $\mathcal{E}(x^\LCm)$ is of infinite duration and infinitely slow variation, $F^\text{ext}$ becomes a constant and uniform electric field, which does obey Maxwell's equations. We will return to this case later.

\subsection{Orbit and impulse}
%
We begin by solving (\ref{LorentzForce}) for the electron orbit. It is trivial to do so in perturbation theory, formally treating $e$ as small. Let the particle have initial momentum $p_\mu$ at initial position $X_\text{in}^\mu$, and define $\pi_\mu(\tau) := m \dot{X}_\mu(\tau)$ the kinematic momentum of the particle (so $\pi^2(\tau) = m^2$). The leading order solution of (\ref{LorentzForce}) in our field is then
\be
    \pi_\mu(\tau)\simeq p_\mu -b(X^\LCm) {\bar n}_\mu + \frac{p_\LCm}{p_\LCp}b(X^\LCm)n_\mu + \ldots\, \quad X^\LCm \simeq X^\LCm_\text{in} + \frac{p_\LCp}{m}\tau + \ldots
\ee
Something is however amiss; writing out the null momentum components $\pi_\LCm$ more explicitly,
\be
    \pi_\LCpm(\tau) \simeq p_\LCpm \bigg( 1 \mp \frac{b(X^\LCm(\tau))}{p_\LCp}\bigg) \;,
\ee
we see that if $p_\LCp$ is of the same order as $b\sim e\mathcal{E}$ then, depending on sign, one of the null momenta can vanish or become negative. This is not allowed for massive particles, which must have $\pi_\LCpm > 0$; only massless particles can have vanishing null momenta. This indicates that the dimensionless parameter controlling perturbation theory goes like $b/p_\LCp$, and that the peturbative expansion breaks down at $ b/p_\LCp\gtrsim 1$, forcing us to resum or use other methods. Fortunately, (\ref{LorentzForce}) is exactly solvable in our chosen field:  for an electron with initial momentum $p_\mu$ the first integral is~\cite{Tomaras:2000ag,Tomaras:2001vs}
\be\label{first-integral}
	\pi_\LCp = p_+ - b(X^\LCm) \;,
 \qquad 
	\pi_\LCm = \frac{m^2 + p_\LCperp^2}{2\left(p_+ - b(X^\LCm) \right)} \;,
 \qquad
	\pi_\LCperp = p_\LCperp \;.
\ee
(The system is thus functionally two-dimensional.) 
Taking the first equation of (\ref{first-integral}), we observe that we can trade proper time $\tau$ for lightfront time $x^\LCm$ via
\be\label{eq:trade}
    \tau = \int^{x^\LCm} \!\!\ud y \, \frac{m }{p_\LCp - b(y)} \;.
\ee
We can thus parameterise the orbit by lightfront time, and integrating once more we obtain the orbit $X^\mu(x^\LCm)$ with components $X^\LCm = x^\LCm$ and 
\be\label{second-integral} \begin{split}
    X^\LCp = \int^{x^{\LCm}}\!\! \ud y \, \frac{m^2 + p_\LCperp^2}{2(p_+ - b(y))^2} \;, \qquad
	X^\LCperp = p^\LCperp\int^{x^{\LCm}}\!\! \ud y \, \frac{1}{p_+ - b(y)} \;.
\end{split} \ee
This exact solution still becomes singular at `critical' times $x^\LCm_c$ solving
\begin{figure}
    \centering
    \includegraphics[width=0.33\textwidth]{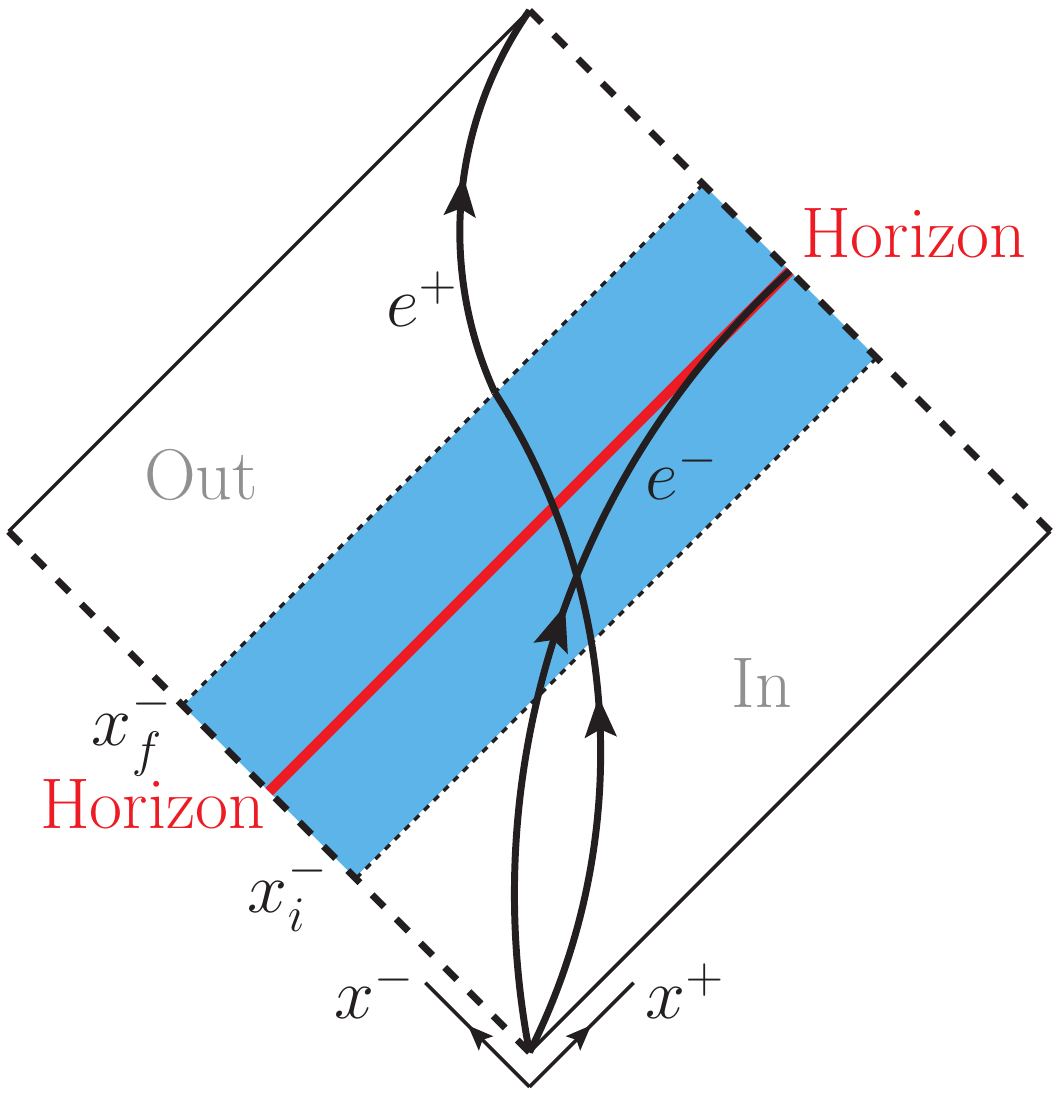}
    \qquad \includegraphics[width=0.4\textwidth]{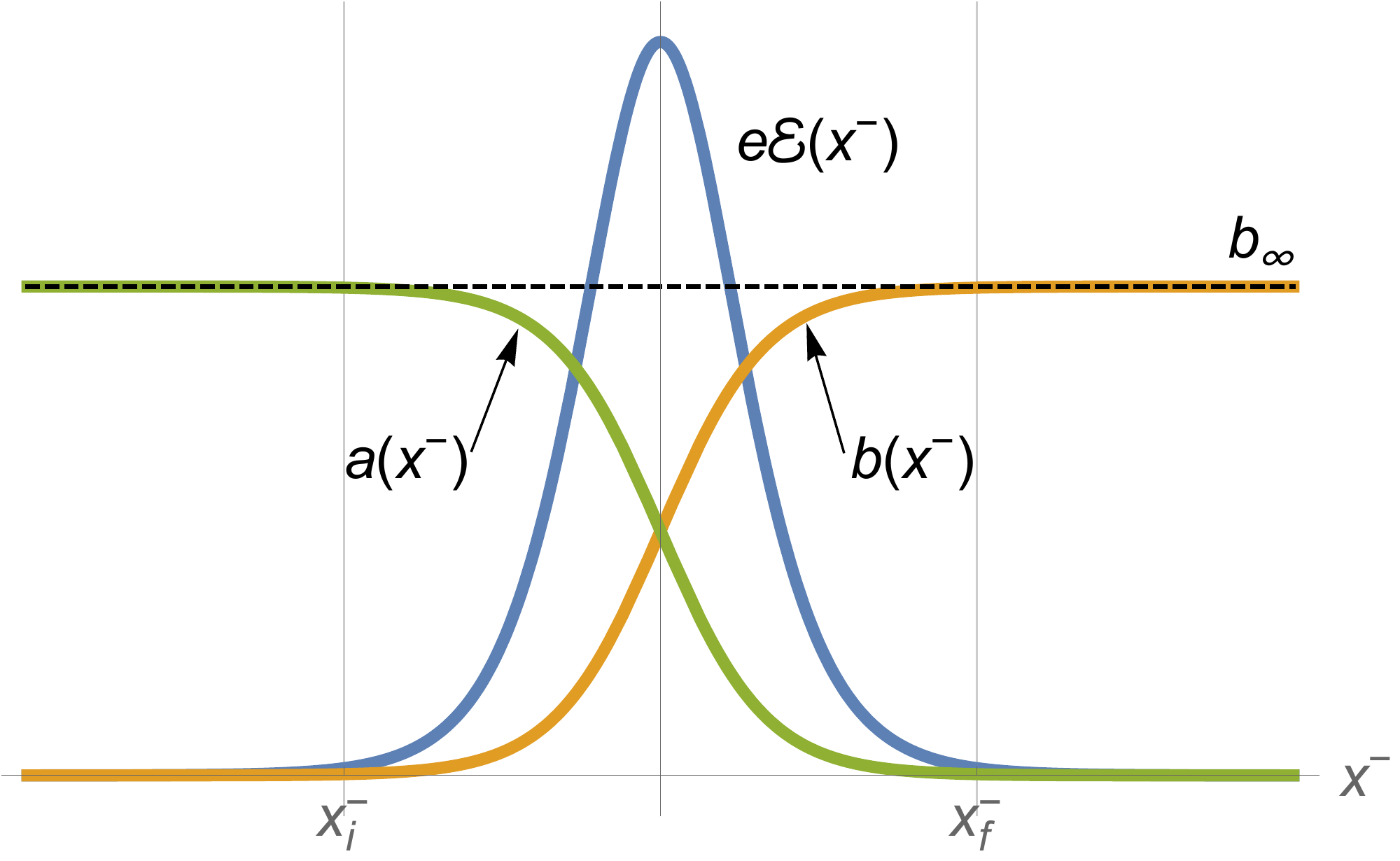}
    \caption{\label{fig:classical}
    \emph{Left:} Our classical system. The sandwich wave divides (Minkowski) spacetime into `in' and `out' regions. Two orbits are shown, one for particles which traverse the field, another for those which are trapped and accelerated by the field toward the speed of light. For such particles the out-region is causally disconnected. \emph{Right:} An illustrative sandwich field profile, taking $e\mathcal{E}>0$, along with its classical potential (\ref{eq:definition-b}) and the associated potential $a(x^\LCm)$ used later in quantum calculations.} 
\end{figure}
%
\be
\label{horizon-condition}
    p_\LCp - b(x^\LCm_c) = 0 \;.
\ee
At this instant $\pi_\LCp\to 0$ and $\pi_\LCm\to \infty$, from (\ref{first-integral}), meaning the electron has been accelerated to the speed of light by the field and, from (\ref{second-integral}), has moved an infinite distance in the $x^\LCp$ (and $x^\LCperp$) directions. The situation is sketched in Fig.~\ref{fig:classical}. If (\ref{horizon-condition}) has a solution, then an incoming electron enters the field but can never leave it, instead being accelerated to the speed of light, and to asymptotically large distance in \emph{finite} lightfront time (but, from (\ref{eq:trade}), infinite proper time). It becomes, in part, causally disconnected from the out-region, and in fact from the larger region $x^\LCm> x^\LCm_c$: no signal from this region can ever be received by the electron. What is happening is that the field (\ref{TheField}) gives a means of generating an acceleration horizon at $x^\LCm=x^\LCm_c$. Indeed, if we take the constant field limit of our results, so $\mathcal{E}(x^\LCm) \to \mathcal{E}$ constant, then the acceleration becomes uniform and the particle orbit is that of a Rindler observer. We therefore refer to (\ref{horizon-condition}) as defining an electromagnetic horizon.

If there are multiple solutions to (\ref{horizon-condition}), only that at lowest $x^\LCm$ is relevant, as this is causally encountered first by particles, and they can never cross the horizon.
We can therefore restrict our attention (until Sect.~\ref{sect:multiple}) to the case $e\Epsilon \ge 0$, i.e.~where the electric field never changes sign, for then $b(x^\LCm)$ is a positive increasing function, and there is at most one solution to (\ref{horizon-condition}): there is a horizon for all electrons with initial momentum obeying $b_\infty/p_\LCp \geq 1$, where 
\be
    b_\infty := \int_{-\infty}^\infty \ud y \, e \Epsilon(y) \;.
\ee
For larger $p_\LCp$ the field cannot capture the electrons, and there is no horizon;  these electrons reach the out-region, but with reduced lightfront momentum $p_\LCp-b_\infty$.

Since opposite charges move in opposite directions in a given electromagnetic field, a region forbidden to electrons can be accessible to positrons. We should therefore expect to be able to establish a horizon for either electrons or positrons, but not both, in contrast to gravity.  The positron orbit is obtained by sending $e\to -e$, so $b\to -b$ in the above. It follows directly that there is no horizon for any positron, as $p_\LCp + b(x^\LCm)$ always remains positive.

Let us summarise the classical situation using (\ref{first-integral}); it will be useful to refer back to this when we consider the quantum theory.
\bi
	\item[1.] Electrons with initial momentum $p_\LCp \in (b_\infty,\infty)$ traverse the field. Their final momentum lies in the range $(0,\infty)$.
	\item[2.] Electrons with momentum $p_\LCp \in (0,b_\infty]$ never pass their horizon.
    \item[3.] All positrons, $p_\LCp \in (0,\infty)$, traverse the field. Their final momentum lies in the range $(b_\infty,\infty)$.
\ei
We note two things. First, there is a minimum final momentum for positrons, something which will be important when we turn to the quantum theory. Second, and already telegraphed, the field exhibits an electromagnetic memory effect~\cite{Bieri:2013hqa}; the momentum of particles which traverse the field is different to the initial momentum. This difference, the \emph{impulse} $\Delta p_\mu$, is 
\be\label{eq:impulse}
    \Delta p_\mu :=\pi_\mu(\infty) - p_\mu  =  - {\bar n}_\mu b_\infty + n_\mu \frac{p_\LCm b_\infty}{p_\LCp - b_\infty}\;,
\ee
for electrons, and $b_\infty\to -b_\infty$ for (all) positrons.

The horizon is not a feature of more commonly-studied electromagnetic backgrounds, see~\cite{DiPiazza:2011tq,Gonoskov:2021hwf,Fedotov:2022ely} for reviews. There is no horizon in plane waves, for example -- all massive particle traverse such fields. Nor is it possible for particles to be trapped in $x^0$--dependent sandwich fields, as is clear on geometrical grounds.

\subsection{Radiation}\label{sect:classical-R}
%
In this and the following subsections we go beyond geodesic motion. We first consider perturbative corrections to the electromagnetic field of the system generated through radiation from the accelerated particles above, and in the next section turn to self-force effects on the particles themselves.

The probe particle considered above has the associated current
\be
    j_\mu(x) = \frac{e}{m}\int\!\ud\tau\,  \pi_\mu(\tau)\, \delta^4(x -X(\tau)) \;, 
\ee
with $\pi_\mu$ as in (\ref{first-integral}) and $X^\mu$ the orbit in (\ref{second-integral}). From this we can calculate the leading-order power emitted by the particle via the standard expression~\cite{Coleman:1961zz}
\be\label{eq:K}
	K_\mu = \int\!\ud k_{\text{o.s.}}\, k_\mu\, j^\dagger_\nu(k) j^\nu(k) \;,
\ee
where $\ud k_{\text{o.s.}}$ stands for the Lorentz-invariant on-shell measure.
The Fourier-transformed current $j_\mu(k)$ is simply
\begin{equation}
    j_\mu(k) = \frac{e}{m}\int\!\ud\tau\, e^{ik\cdot X(\tau)} \pi_\mu(\tau) = e\int_{-\infty}^{x^\LCm_c}\!\ud x^\LCm\, e^{ik\cdot X} \frac{\pi_\mu(x^\LCm)}{p_\LCp - b(x^\LCm)},
\end{equation}
where we have used (\ref{eq:trade}) to convert from proper time to lightfront time, and where $x^\LCm_c = b^{-1}(p_\LCp)$ if the particle cannot cross the horizon, but $x^\LCm_c = \infty$ if it can escape the field. The integrals in (\ref{eq:K}) can now be performed and are elementary. The result is
\be\label{eq:K-explicit}
	K_\mu = \frac23 \frac{e^2}{4\pi} \frac{m^2+p_\LCperp^2}{m^4}\int_{-\infty}^{x_c^\LCm} \!\ud x\, \frac{\mathcal{E}(x)^2}{p_\LCp- b(x)} \pi_\mu(x) \;.
\ee
To analyse this expression we set $p_\LCperp=0$ for simplicity, so $K_\LCperp=0$. If there is no horizon, both $K_\LCp$ and $K_\LCm$ are finite and nonzero. If there is a horizon, so $x_c^\LCm = b^{-1}(p_\LCp)$ is finite, then $K_\LCp$ remains finite while $K_\LCm$ diverges. This is because the particle acceleration asymptotes, over infinite proper time, to the $n_\mu$ direction, and since high energy particles emit roughly forward, the radiated momentum is dominated by its $n_\mu \sim \delta_\mu^\LCm$ component.

\subsection{All-orders self-force}\label{sect:RR}
A natural question to ask is self-force effects (i.e.~radiation, above, and radiation reaction) change the essential picture; can a radiating electron cross the horizon? Does radiation lead to an effective horizon also for positrons?

Quantitative answers are provided by solving the Landau-Lifshitz equation for the trajectory of a \emph{radiating} particle -- this is the equation obtained by applying reduction of order to the Lorentz-Abraham-Dirac equation in order to remove its unphysical runaway and pre-accelerating solutions~\cite{LL}: the equation is, writing $f_{\mu\nu}:=e F^\text{ext}_{\mu\nu}/m$ to compactify notation, 
\be\label{eq:LL}
    \ddot{X}_\mu = f_{\mu\nu}(X){\dot X}^\nu
    + \frac23 \frac{e^2}{4\pi m}
    \bigg(
    {\dot f}_{\mu\nu}(X) 
    + f^2_{\mu\sigma}(X){\dot X}^\sigma {\dot X}_\nu
    - {\dot X}_\mu f^2_{\nu\sigma}(X){\dot X}^\sigma
    \bigg){\dot X}^\nu \;.
\ee
The gravitational analogue of (\ref{eq:LL}) is the reduced-order MiSaTaQuWa equation, as described in~\cite{Quinn:1996am}. The Landau-Lifshitz equation was solved for our system in~\cite{Ekman:2021vwg}, so we present only the key steps here. First, since relevant motion is only in the $(x^\LCp,x^\LCm)$ plane, we look for a solution with ${\dot{X}}^\LCperp=0$ (see~\cite{Ekman:2021vwg} for the extension), which means the mass-shell condition reduces to $2{\dot X}^\LCp \dot{X^\LCm}=1$.  With this, dotting $n_\mu$ into (\ref{eq:LL}) yields
\[
    \ddot{X}^\LCm = -e\mathcal{E}(X^\LCm){\dot X}^\LCm  - \frac23 \frac{e^2}{4\pi m} \mathcal{E}'(X^\LCm){\dot X}^\LCm {\dot X}^\LCm \;.
\]
Changing variable from proper time to lightfront time, this equation is easily integrated to find the solution for $\pi_\LCp = m {\dot X}^\LCm$:
\be
    \pi_\LCp(x^\LCm) = \mathrm{e}^{-e^3\mathcal{E}(y)/(6\pi m^2)} \bigg[ p_\LCp - \int_{-\infty}^{x^\LCm}\!\ud y\, \mathrm{e}^{e^3\mathcal{E}(y)/(6\pi m^2)} e\mathcal{E}(y)\bigg] \;,
\ee
which reduces to the solution (\ref{first-integral}) of the Lorentz force equation in the limit that we neglect radiation reaction, so $e^2\to 0$ at fixed $e\mathcal{E}$. We thus read off that the horizon is now located at the solution of
\be\label{RR}
    p_\LCp - \int_{-\infty}^{x^\LCm}\!\ud y\, \mathrm{e}^{e^3\mathcal{E}(y)/(6\pi m^2)} e\mathcal{E}(y) =0 \;,
\ee
which is explicitly all-orders in the coupling. Because the exponent in (\ref{RR}) is positive, it means that the horizon remains and is shifted to `earlier' times $x^\LCm < b^{-1}(p_\LCp)$. In essence, the electron radiates, and loses energy, thus is more easily trapped by the field. The effective value of $x^\LCm$ at which the horizon lies is reduced, and more of the spacetime is excluded. It is also clear from swapping the sign of $e\mathcal{E}$ in (\ref{RR}) that there remains no horizon for positrons.

\section{Quantum} \label{sect:quantum}
%
In this section we want to explain how the classical horizon found above appears in quantum calculations. To do so we will introduce the wavefunctions which describe massive particles in the background (\ref{TheField}) and are the input for all scattering calculations. We will see that the wavefunctions know about the horizon through their spacetime singularities. We will then interpret their behaviour at the horizon as a signal of pair production via the Schwinger effect, and compare to Hawking radiation.

The theory of interest is scalar QED, with the additional background field (\ref{TheField}). 
We will give a brief outline of how we calculate amplitudes in this theory. (Details will come later, along with the extension to spinor QED.) Defining the background-covariant derivative
\be
    \mathcal{D}_\mu := \partial_\mu + i e A^{\text{ext}}_\mu \;,
\ee
the action is
\be\label{scalar-action}
    \int\!\ud^4x\, \bigg[-\frac14 F^{\mu\nu}F_{\mu\nu}
    + (\mathcal{D}_\mu\phi)^\dagger \mathcal{D}^\mu\phi -m^2 \phi^\dagger \phi
    + i e A_\mu \big[(\mathcal{D}^\mu\phi)^\dagger \phi - \phi^\dagger \mathcal{D}^\mu\phi \big]  +e^2 A^2 \phi^\dagger \phi \bigg]      \;,
\ee
in which $F$ is the fieldstrength of $A$.
Mirroring the classical discussion, we begin by neglecting radiation and radiation reaction. This means dropping the interaction terms going like $eA_\mu$ and $e^2A^2$ in (\ref{scalar-action}),  so that no photons can be generated or absorbed.  We are thus left with a massive scalar field in a fixed background. The equation of motion for the field is the Klein-Gordon (KG) equation in the background, and thus all observables are built from the \emph{solutions of those equations.} For example, two-point amplitudes in this truncated theory are simply appropriate overlaps of the solutions, and encode the quantum analogue of the classical impulse computed via the Lorentz force equation (\ref{LorentzForce}). To compute the exact impulse, as we did classically, we must 
treat the coupling to the background $A^\text{ext}$ exactly, not perturbatively. (In our case we already know this is dimensionless $b_\infty/p_\LCp$.)

To reintroduce radiation (reaction), we will later turn the $eA$ and $e^2A^2$ terms back on in (\ref{scalar-action}), treating them in perturbation theory. This generates the usual diagrammatic expansion of scalar QED except that massive internal lines represent the exact KG propagator in the background, and massive external lines are solutions of the KG equation. This is the Furry expansion, or background field perturbation theory~\cite{Furry:1951zz,DeWitt:1967ub,tHooft:1975uxh,Abbott:1981ke}. It is the analogue of the self-force expansion in gravity, about a given metric. So, whether we include radiation reaction or not, we need the exact solutions of the KG equation in our background. We compute and explore them now, beginning with the scalar case.

\subsection{Wavefunctions}
We want to construct incoming and outgoing solutions of the KG equation which reduce to free states in the asymptotic past and future. To clarify this, observe that in our system we cannot specify free data at fixed $x^0$, as such surfaces always intersect the background. Instead we can use a foliation of hypersurfaces of constant lightfront time $x^\LCm$, which is natural given the geometry of our problem. Our boundary conditions are then that wavefunctions reduce to free wavefunctions $e^{\pm ip\cdot x}$ as $x^\LCm\to \pm\infty$.   These null surfaces are not, note, Cauchy surfaces, but rather characteristics, and this will have consequences.

For comparison, there are clear similarities to the case of plane wave backgrounds in both electrodynamics and gravity. A sandwich plane wave is, like our field, a compactly supported function of $x^\LCm$ which similarly separates spacetime into `in' and `out' regions at surfaces of constant $x^\LCm$.
Spacetime is not asymptotically flat in the null direction of the plane wave and there is again no Cauchy surface~\cite{Penrose:1965rx}, but there is a well-defined, and well-studied, $S$-matrix~\cite{Penrose:1965rx,Neville:1971uc,Gibbons:1975jb,Garriga:1990dp,Adamo:2017nia} -- it is here that we will find differences with our system.

\subsubsection*{Incoming}
The exact solutions of the KG (or conjugate KG) equation with incoming  are
\be
\label{electron-in}
\text{electrons:}	\quad e^\LCm_{\text{in}}(p;x) = e^{-ip\cdot x + i b(x^\LCm)x^\LCp}\sqrt{\frac{p_\LCp}{p_\LCp- b(x^\LCm)}} \exp \bigg[ - i \int_{-\infty}^{x^\LCm}\!\ud z\, \frac{p_\LCm b(z)}{p_\LCp -b(z)}\bigg] \;,
\ee
\be
\label{positron in}
\text{positrons:} \quad e^\LCp_{\text{in}}(p;x)  = e^{-ip\cdot x - i b(x^\LCm)x^\LCp}\sqrt{\frac{p_\LCp}{p_\LCp+ b(x^\LCm)}} \exp \bigg[ + i \int_{-\infty}^{x^\LCm}\!\ud z\, \frac{p_\LCm b(z)}{p_\LCp +b(z)}\bigg] \;.
\ee
(These are related simply by swapping the sign of the charge, so $b\to-b$.)
For all positrons, and for electrons with initial momentum $p_\LCp \in (b_\infty,\infty)$ the wavefunctions are everywhere well-behaved. They evolve smoothly from the in-region to the out-region, just as classical particles with the same initial momentum would traverse the field. In this case, (\ref{electron-in}) and (\ref{positron in}) are just single-particle functions which, in the asymptotic regions, are free wavefunctions obeying
\be\label{eq:memory}
	\text{in:}\quad \frac{1}{\sqrt{2p_\LCp}} \, e^{-ip\cdot x}
    \quad\longrightarrow\,
    \quad
       \text{out:}\quad
       \frac{1}{\sqrt{2{\pi_\LCp(\infty)}}} \,
    e^{-i{ \pi(\infty)}\cdot x}e^{-i\theta}
\ee
where $\theta$ is a finite, time-independent phase and ${\pi}_\mu(\infty)$ is the asymptotic classical momentum, as in (\ref{eq:impulse}). The normalisation factor is included here to demonstrate the role of the square-root prefactors in (\ref{electron-in})--(\ref{positron in}): they ensure the wavefunctions stay correctly relativistically normalised as the particle momentum changes. The wavefunctions thus  exhibit the electromagnetic memory effect we saw in the classical theory. 

Indeed it is interesting to note that if we reinstate $\hbar$ we can see that these wavefunctions are almost semiclassical-exact; the exponent is just the classical Hamilton-Jacobi action over $\hbar$, i.e.~the leading order WKB solution of the KG equation, while the square root (a logarithm in disguise) is the next-to-leading order, $\hbar$-independent WKB term.

The situation for electrons with $p_\LCp \in (0,b_\infty]$ is different; in this case the wavefunctions (\ref{electron-in}) are singular. They diverge at the classical horizon, $p_\LCp - b(x^\LCm)=0$, as in (\ref{horizon-condition}).  The wavefunctions, and hence quantities constructed from them, thus know about the horizon. (See e.g.~\cite{Fabbrichesi:1993kz,Parnachev:2020zbr} for how eikonal amplitudes are sensitive to horizons in AdS, and~\cite{Ferrero:2021lhd} for how amplitudes in de Sitter encode the causal structure of the spacetime.) To compute amplitudes, we will have to answer the question of how to define the wavefunctions at, and across, the horizon. A similar question arises for outgoing wavefunctions, as we will now see.

\subsubsection*{Outgoing}
We introduce, in addition to $b(x^\LCm)$, the complementary integral
\be
	a(x^\LCm) := \int^{\infty}_{x^\LCm} \!\ud s \, e\mathcal{E}(s) \;,
\ee
which vanishes in the asymptotic future. As such outgoing wavefunctions are naturally expressed in terms of $a$ rather than $b$, the relation between the two being
\be
a(x^\LCm) + b(x^\LCm) = \int_{-\infty}^{\infty}\!\ud s\, e\mathcal{E}(s) = b_\infty\,.
 \ee
The outgoing solutions of the (conjugate KG or) KG equation are:
\be
\label{electron-out}
	\text{electrons:}\quad
    e^\LCm_\text{out}(p;x) = 
	\displaystyle e^{ip\cdot x + i a(x^\LCm)x^\LCp}\sqrt{\frac{p_\LCp}{p_\LCp + a(x^\LCm)}} \exp \bigg[ + i \int^{\infty}_{x^\LCm}\!\ud z\, \frac{p_\LCm a(z)}{p_\LCp +a(z)}\bigg] \;,
\ee
\be
\label{positron-out}
	\text{positrons:}\quad
    e^\LCp_\text{out}(p;x) = 
	\displaystyle e^{ip\cdot x - i a(x^\LCm)x^\LCp}\sqrt{\frac{p_\LCp}{p_\LCp- a(x^\LCm)}} \exp \bigg[ - i \int^{\infty}_{x^\LCm}\!\ud z\, \frac{p_\LCm a(z)}{p_\LCp -a(z)}\bigg] \;.
\ee
The outgoing electron wavefunctions are now nonsingular everywhere. They can be traced back to the in-region, where they describe free particles with incoming momentum in the range $p^\text{in}_\LCp \in (b_\infty,\infty)$.  This is consistent with the classical result that only electrons with momentum in this range cross the horizon to the out-region. Similarly, outgoing positron wavefunctions with $p_\LCp \in (b_\infty,\infty)$ are nonsingular. Traced back to the in-region they describe free particles with initial momenta in the range $(0,\infty)$. This is again consistent with the results that (i) all incoming classical positrons traverse the field and have final momentum {in the range} $(b_\infty,\infty)$, and (ii) all incoming positron wavefunctions are nonsingular. The kinematic limits imposed by the horizon are thus visible even in completely well-behaved wavefunctions. 

It remains to consider positrons (\ref{positron-out}) with $p_\LCp<b_\infty$. Classically, we know it is not possible for these positrons to have traversed the field; all incoming positrons cross the horizon, but they then have $p_\LCp > b_\infty$. What, then, do these solutions mean? Tracing the wavefunctions back in time, we see that they become singular at
\be\label{out-horizon}
    p_\LCp - a(x^\LCm) = 0 \;.
\ee
This is a past horizon for outgoing positrons. We now have to answer the questions of how to continue the wavefunctions across the horizon, and what these solutions mean physically.

\subsection{Crossing the horizon, losing unitarity, creating pairs}
\label{sect:unitarity}
To continue our analysis of the wavefunctions above we now ask what happens at the horizon.

Consider first the incoming electron wavefunction (\ref{electron-in}) for $p_\LCp < b_\infty$. As we approach the horizon at $b^{-1}(p_\LCp)$ we continue the integral in the exponent into the upper half plane and go around the pole\footnote{This is a choice; it is the physically correct choice as we will see, but a choice nonetheless. We address this later in the text.}. In doing so the integral acquires an imaginary part, hence the exponent acquires a \emph{real} part, and the wavefunction becomes
\be
\label{continued}
e^{-ip\cdot x + i b(x^\LCm)x^\LCp}\sqrt{\frac{p_\LCp}{b(x^\LCm)-p_\LCp}} e^{-i\pi/2}\exp \bigg[ - i \text{P.V.}\int_{-\infty}^{x^\LCm}\!\ud z\, \frac{p_\LCm b(z)}{p_\LCp -b(z)}\bigg]\exp\bigg[ - \frac{\pi p_\LCp p_\LCm}{e\mathcal{E}(b^{-1}(p_\LCp))}\bigg] \;,
\ee
The wavefunction now extends into the out-region.
(The additional phase comes from evaluating the square root at negative argument when $x^\LCm > b^{-1}(p_\LCp)$.) 
We might expect that quantum particles can pass the horizon even if classical particles cannot, suggesting  that the final, real exponential in (\ref{continued}) is a tunnelling exponent. However, the wavefunction has no overlap with \emph{outgoing} states: it does not describe a particle which has tunneled through the field. To see this, we only need to observe (see Sec.~\ref{sect:scattering}
for details) that any overlap between (\ref{continued}) and an outgoing state of momentum $q_\mu$, evaluated in the asymptotic future, i.e.~the out region, will include an integral like
\begin{equation}
\label{eq:new}
    \int\!\ud x^\LCp e^{i q_\LCp x^\LCp} e^{-i p_\LCp x^\LCp + i b(x^\LCm)}  \sim \delta(q -p + b(x^\LCm)) = 0 \;,
\end{equation}
which vanishes because both $q_\LCp$ and $b-p$ are positive beyond the horizon.
Another problem is that, because of the real exponential, the norm of the corresponding state is not preserved as we cross the horizon.
It seems that unitarity is violated. This is emphasised by the observation that there are other solutions to the KG equation, described in Appendix~\ref{app:zeromodes}, which switch on or off at the horizon: any of these can be added to our wavefunctions at and beyond the horizon, but there is nothing to dictate which or why.
This is clearly at odds with the idea of a uniquely defined $S$-matrix.

The situation is similar for the outgoing positron wavefunction (\ref{positron-out}) for $p_\LCp < b_\infty$.  As we approach the horizon (\ref{out-horizon}) at $a^{-1}(p_\LCp)$, we again go around the pole in the upper half plane, which generates an exponential damping term as before:
\be\label{continued2}
	e^{ip\cdot x - i a(x^\LCm)x^\LCp}\sqrt{\frac{p_\LCp}{a(x^\LCm) - p_\LCp}} e^{-i\pi/2}\exp \bigg[- i \text{P.V.}\int^{\infty}_{x^\LCm}\!\ud z\, \frac{p_\LCm a(z)}{p_\LCp -a(z)}\bigg]\exp\bigg[-\frac{\pi {p_\LCm p_\LCp} }{\mathcal{E}(a^{-1}(p_\LCp))}\bigg] \;,
\ee
and the wavefunction now extends beyond the past horizon.

What is the interpretation of (\ref{continued2})? First, we know that these positron wavefunctions cannot be related to purely classical physics, since all positrons which traverse the field have a final momentum greater than $b_\infty$.
Second, it is known that the field (\ref{TheField}) exhibits the Schwinger effect, i.e.~there is spontaneous pair production~\cite{Tomaras:2000ag,Tomaras:2001vs}.
Indeed we recognise in (\ref{continued2}) the Schwinger-type exponent typical of nonperturbative pair production. We therefore interpret the existence and behaviour of (\ref{continued2}) in terms of pair production. In particular, (\ref{continued2}) describes positrons \emph{produced} by the field. Curiously, and independent of how we continue them, the positron wavefunctions (\ref{positron-out}) are clearly \emph{single}-particle wavefunctions in the out-region, and none of our wavefunctions describe electrons which were not already \emph{initially} present. The conclusion, following~\cite{Tomaras:2000ag,Tomaras:2001vs}, is not that charge conservation is violated, but that the created electron is itself trapped by the field, and never escapes to the out-region. What we demonstrate in the next section is that the discontinuity of the positron wavefunctions (\ref{continued2}) at the horizon contains information on pair production.  This is consistent with the literature on the Hawking effect, in which the discontinuity of wavefunctions, analytically continued across the black hole horizon, plays an essential role~\cite{Hawking:1975vcx,Damour:1976jd}. (For a recent comparison and review of the methods used therein see~\cite{Li:2016sjq}.)

\subsection{Cauchy vs. characteristic data}\label{sec:CC}
Before continuing, we very briefly mention some alternative approaches to the problems encountered above. The system we consider was analysed in~\cite{Tomaras:2000ag,Tomaras:2001vs}. There, the background (\ref{TheField}) was truncated to a semi-infinite range in $x^\LCp$. This allows one to specify free data on \emph{both} characteristic surfaces $x^\LCm=$ constant and $x^\LCp=L$, which is equivalent to specifying \emph{Cauchy} data. Evolving the fields in a two-time formalism from both surfaces~\cite{Bogolyubov:1990kw,Heinzl:1991vd,Tomaras:2000ag,Tomaras:2001vs}, the extra boundary data dictates both the pole prescription at the horizon and the presence of additional terms in the wavefunctions, which switch on at the horizons and maintain unitarity. We give a more complete discussion in Appendix~\ref{app:TTW}. A two-time formalism is somewhat unintuitive, though, so here we continue to pursue an amplitudes-based approach. In the next section we will use the perturbiner method to obtain amplitudes in our background, and demonstrate that this correctly computes pair production observables.

\section{Amplitudes and horizons}
\label{sect:scattering}
%
%
In the perturbiner approach to scattering, an $n$-point tree-level amplitude is computed by the $n$-linear piece of the classical action, evaluated on particular solutions of the classical equations of motion~\cite{Arefeva:1974jv,Abbott:1983zw,Jevicki:1987ax}, chosen to impose the desired incoming/outgoing boundary conditions on the scatterers. When the theory in question admits an $S$-matrix, the perturbiner approach recovers amplitudes calculated by standard means. The power of the approach is that it provides, even when the $S$-matrix does not exist, a notion of scattering amplitudes as objects encoding the information needed to construct observables. 

We begin with 2-point amplitudes. These are not trivial on backgrounds, as $1\to 1$ amplitudes encode geodesic motion and thus memory effects, while $0\to 2$ amplitudes encode pair production\footnote{Even in vacuum, we note that 2-point string amplitudes continue to be a source of new results, see~\cite{Erbin:2019uiz,Giribet:2023gub}.}.  The 2-point amplitude in the perturbiner approach is, in general, a boundary term. This follows directly from taking the quadratic part of the action,
\be\label{eq:quadratic}
      \mathcal{S}_2[\phi] =   \int\!\ud^4 x\, (\mathcal{D}_\mu\phi)^\dagger \mathcal{D}^\mu\phi - m^2\phi^\dagger\phi\;,
\ee
and using integration by parts:
\be\label{eq:quadratic2}
      \mathcal{S}_2[\phi] = \int\!\ud S^\mu \, \phi^\dagger {\mathcal{D}}_\mu\phi - \cancel{\int\!\ud^4x \, \phi^\dagger \big({\mathcal{D}}^2+m^2\big)\phi}\,.
\ee
The second term vanishes when $\phi$ is a solution of the KG equation on the background, while $\ud S^\mu$ implicitly parameterises \emph{all} boundary surfaces, the set of which depends on the system. For example, if the field is asymptotically flat the set contains the (conformally compactified) boundary of Minkowski space. Evaluated in asymptotically flat regions, (\ref{eq:quadratic2}) immediately reduces to the usual free-field inner product, and as such coincides with textbook LSZ expressions. As another example, a shockwave background has two additional boundaries, on either side of the discontinuity defined by the shock~\cite{tHooft:1987vrq,Jackiw:1991ck,Fabbrichesi:1993kz,Adamo:2021jxz,Adamo:2021rfq}. For a Schwarzschild background there are boundaries at spatial infinity, and the event horizon, and the $S$-matrix does not exist.  However, there are physical scenarios which should not require us to deal with the subtleties of the event horizon~\cite{Giddings:2007qq,Adamo:2022rob}, an example being scattering at large distance. In this case the perturbiner approach computes amplitudes by evaluating the gravitational analogue of (\ref{eq:quadratic2}) on \emph{only} the boundary at infinity, see~\cite{Adamo:2021rfq}, thus providing access to observables even though there is no S-matrix.

Perturbiner methods also underlie some well-known calculations: for example, the computation of the 2-point amplitude in AdS which, when expressed in terms of boundary data, exhibits the structure of the boundary CFT~\cite{Witten:1998qj}. Further examples can be found in e.g.~\cite{Rosly:1996vr,Rosly:1997ap} for helicity amplitudes, \cite{Adamo:2017nia} for scattering on plane wave backgrounds, and~\cite{Gonzo:2022tjm} for applications in celestial holography. We will make use of the versatility of the perturbiner approach to calculate the pair creation amplitude in our system. 

\subsection{Pair production}
Consider a superposition $\phi(x) = \epsilon_1 \phi_1(x) + \epsilon_2 \phi_2(x)$ of solutions $\phi_j$ of the KG equation, in which the $\epsilon_j$ are complex parameters. Then $\mathcal{S}_2[\phi]$ encodes several possible 2-point amplitudes which are extracted via e.g.~ 
\be\label{eq:2-point-arb}
    \mathcal{A}_2 = i \frac{\partial^2 {\mathcal S}_2[\phi]}{\partial {\epsilon}_1\partial {\bar\epsilon}_2} \bigg|_{\epsilon_j=0}  = \int\!\ud S^\mu \, \phi_2^\dagger(x) i\overset{\leftrightarrow}{\partial}_\mu \phi_1(x) \;.
\ee
For pair production, we choose both of the $\phi_j$ to be outgoing wavefunctions. There are two non-trivial surfaces to consider. First, the surface $x^\LCm = x_f^\LCm$ which separates the field from the out-region. A direct calculation of (\ref{eq:2-point-arb}) on this boundary with the appropriate combination of outgoing wavefunctions (\ref{electron-out}) and (\ref{positron-out}) yields a contribution proportional to
\be\label{eq:wrong}
    \int\!\ud x^\LCp\, e^{i(q+p)_\LCp x^\LCp} = 2\pi {\delta}(q_\LCp + p_\LCp) = 0 \;,
\ee
vanishing because of the positivity of the lightfront momenta\footnote{Choosing any parallel surface $x^\LCm = T > x_f^\LCm$ gives the same result, since (\ref{eq:quadratic2}) is just the free KG inner product when the background field vanishes, and is therefore time-translation invariant by definition.}, much as in (\ref{eq:new}). This cannot be the final result, though, since we know there is pair creation in this system. Rather, this calculation reflects the result that only the created positron escapes the field: the electron never reaches the out-region, so the amplitude to find it in the out-region is zero. 

The only remaining boundary is that at $x^\LCm = x^\LCm_i$ separating the field and the in-region. Consider the term $i(p_\LCp-a(x^\LCm))x^\LCp$ in the exponent of the outgoing positron wavefunction (\ref{positron-out}). As we pass the horizon, $p_\LCp-a(x^\LCm)$ flips sign; because this is a lightfront momentum component, it means our positive energy/outgoing solution becomes a negative energy solution `behind' the past horizon and hence also on the surface $x^\LCm = x^\LCm_i$. This implies the wavefunction can, here, have overlap with other outgoing states.

In this way (\ref{positron-out}) encodes information about pair creation, and the created electron, even though it is a single-particle wavefunction in the out-region. This information is `hidden' behind the horizon (\ref{out-horizon}), but we can access it by taking overlaps with a basis of outgoing electron states on the boundary $x^\LCm = x^\LCm_i$. Thus the scalar QED amplitude we need to calculate is
\be\label{eq:todosQED}
\mathcal{A}_\text{SQED} = \int \ud^2 x^\LCperp \ud x^\LCp \, e^{i q \cdot x} \, i\overset{\leftrightarrow}{\partial}_\LCp e^\LCp_\text{out}(x) \bigg|_{x^\LCm = x^\LCm_i} \;,
\ee
which we proceed with immediately. The spatial integrals in (\ref{eq:todosQED}) yield three delta-functions,
\begin{equation}
\label{amplitudes}
	\begin{split}
		\mathcal{A}_\text{SQED} &= 2 i \sqrt{q_\LCp p_\LCp}\,
        \deltahat^2(q_\LCperp + p_\LCperp)
        \deltahat(p_\LCp + q_\LCp - b_\infty)
        e^{i\theta}
        \exp \bigg(- \pi \frac{p_\LCm p_\LCp}{e\Epsilon (x_c)}\bigg) \;,
	\end{split}
\end{equation}
where $x_c^\LCm = a^{-1}(p_\LCp)$ is the location of the horizon, we write $\hat\delta(\cdot) \equiv 2\pi \delta(\cdot)$ and $\theta$, real, is
\be
    \theta = (q_\LCm + p_\LCm)x_i^\LCm + \mathcal{P} \int_{\infty}^{x_i^\LCm} \!\ud y\, \frac{p_\LCm a(y)}{p_\LCp - a(y)} \;.
\ee
Note that the delta function in (\ref{amplitudes}) now sets $p_\LCp+q_\LCp = b_\infty$, rather than zero when evaluating on the boundary $x^\LCm = x^\LCm_f$, see (\ref{eq:wrong}). The field thus allows for the creation of positrons with momenta \emph{only} in the range $0<p_\LCp<b_\infty$; this is in agreement with our earlier findings, as these are the positrons which are classically forbidden from the out-region. Squaring up, we find
\begin{equation}
\begin{split}
   |\mathcal{A}_\text{SQED}|^2 &= 4 V_\text{LF} \,
    q_\LCp (b_\infty-p_\LCp)
    \deltahat^2_\LCperp(q + p)
    \deltahat(q_\LCp+p_\LCp-b_\infty) \exp  \bigg[ - \pi \frac{m^2 + p_\LCperp^2}{e\Epsilon(x^\LCm_c)} \bigg] \;,
\end{split}
\end{equation}
in which $V_{\text{LF}}$ is the three-volume of the $x^\LCp$, $x^\LCperp$ directions which results from squaring the delta functions. The total pair production probability is obtained by integrating over the on-shell momenta of the pair:
\begin{equation}
\begin{split}
    \mathbb{P}_\text{SQED} &= \int\!\ud p_\text{o.s.}\ud q_\text{o.s.}\,  |\mathcal{A}_\text{SQED}|^2 = V_\text{LF} \int\! \frac{\ud^2 p_\LCperp}{(2\pi)^2} \int_0^{b_\infty}\! \frac{\ud p_\LCp}{2\pi} \, \exp \left[ - \pi \frac{m^2 + p_\LCperp^2}{e\Epsilon(x_c)} \right] \;, 
\end{split}
\end{equation}
in which the bounds on the remaining $p_\LCp$ integral result from the delta functions and the positivity of $p_\LCp$ and $q_\LCp$. The integrals over $p_\LCperp$ are Gaussian and can be performed immediately.  For the $p_\LCp$ integral, we note $x^\LCm _c = a^{-1}(p_\LCp)$ is a function of $p_\LCp$, so we may change variable from an integral over momenta to an integral over the locations of the horizon. Doing so, our final expression for the probability of creating a pair is
\begin{equation}
\label{eq:final}
\begin{split}
    \mathbb{P}_\text{SQED} &= \frac{V_\text{LF}}{\left(2\pi\right)^3}
    \int_{x_i^\LCm}^{x^\LCm_f} \!\ud y^\LCm \,\,
    e^2 \Epsilon^2(y^\LCm) \exp \bigg[-\frac{\pi m^2}{e\Epsilon(y^\LCm)}  \bigg] \;.
\end{split}
\end{equation}
This agrees with previous literature results~\cite{Tomaras:2000ag,Tomaras:2001vs,Fried:2001ur,Ilderton:2014mla}. We have thus obtained the pair production probability from what appeared initially to be a single-particle wavefunction. 

Pair production in our system thus exhibits concretely some features popularly ascribed to Hawking radiation.
First, pair production happens precisely at the horizon -- it is an instantaneous event for any given particle momentum, occurring at the singularity of the wavefunctions.
Second, only particles of one charge can escape the horizon, while the opposite charges are trapped by the field.
This is not the physical picture in gravity, but it is here, nor is this behaviour a general feature of pair production in other fields, see~\cite{Dunne:2004nc,Fedotov:2022ely} for reviews.
(What we have calculated is not, strictly, the probability of \emph{observing} a pair, since it is not possible for both particles to escape to the out-region. Rather we have calculated the equivalent probability of observing a positron that was pair-created.)

\subsection{Horizons, quantum interference, and memory}\label{sect:multiple}
The result (\ref{eq:final}) can be obtained from the famous Schwinger formula for pair creation in a constant field $\mathcal{E}$~\cite{Schwinger:1951nm} by replacing $\mathcal{E} \to \mathcal{E}(y^\LCm)$, and a volume factor by the $y^\LCm$--integral. This `locally constant field approximation' to pair production is widely used phenomenologically~\cite{Bulanov:2010ei}, but the result is exact here. This simplicity is particular to $x^\LCm$-dependent fields (see~\cite{Ilderton:2015qda} for an explanation of why, in terms of the underlying analytic structure.)
It seems to be an open question in the literature as to whether this locally constant form also holds for fields which can, unlike the field studied so far, change sign.
We will briefly investigate these points here.

\begin{figure}[t!]
    \centering
    \includegraphics[width=0.34\textwidth]{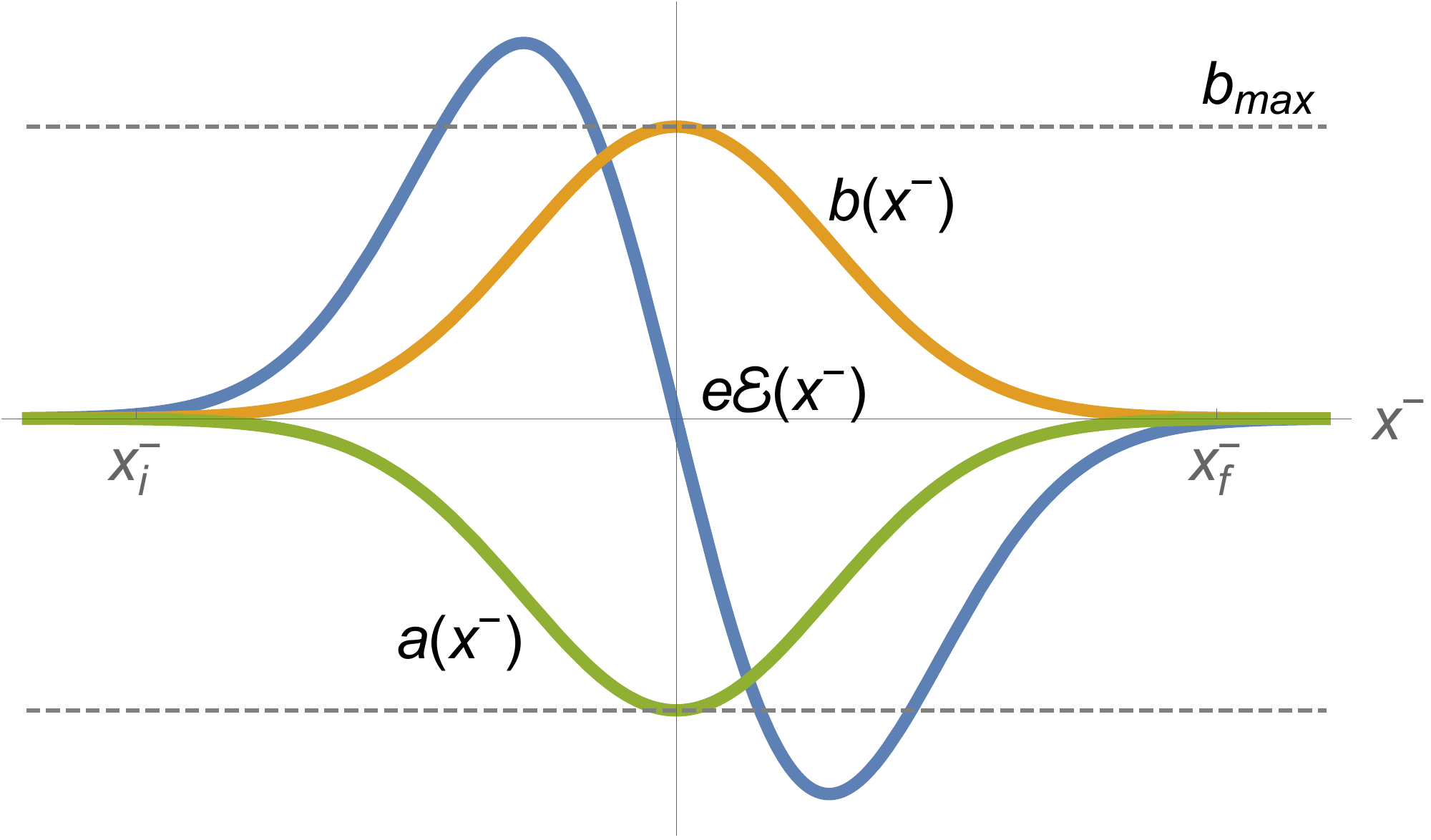}
    \quad
    \includegraphics[width=0.34\textwidth]{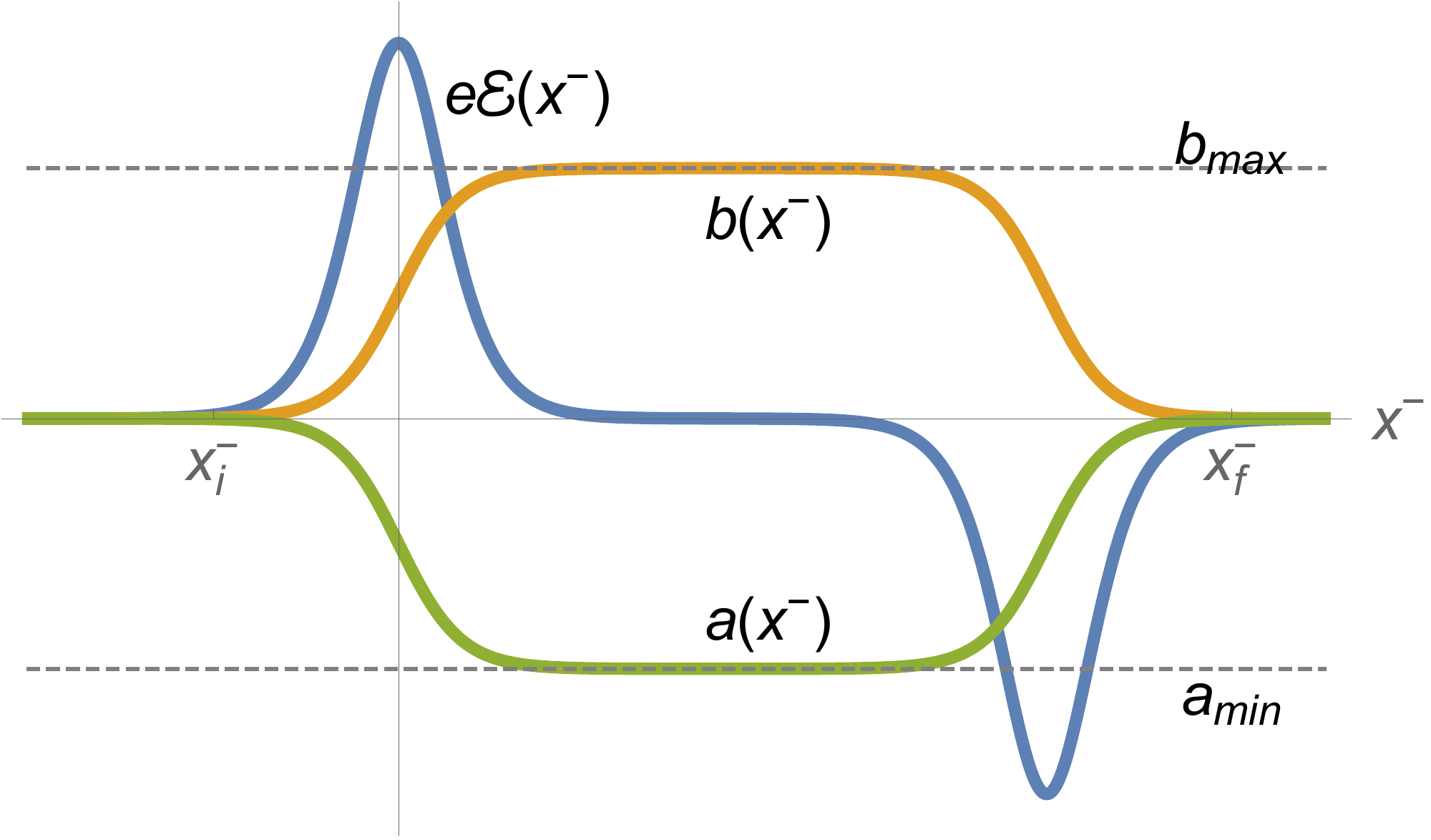}
    \quad
    \includegraphics[width=0.22\textwidth]{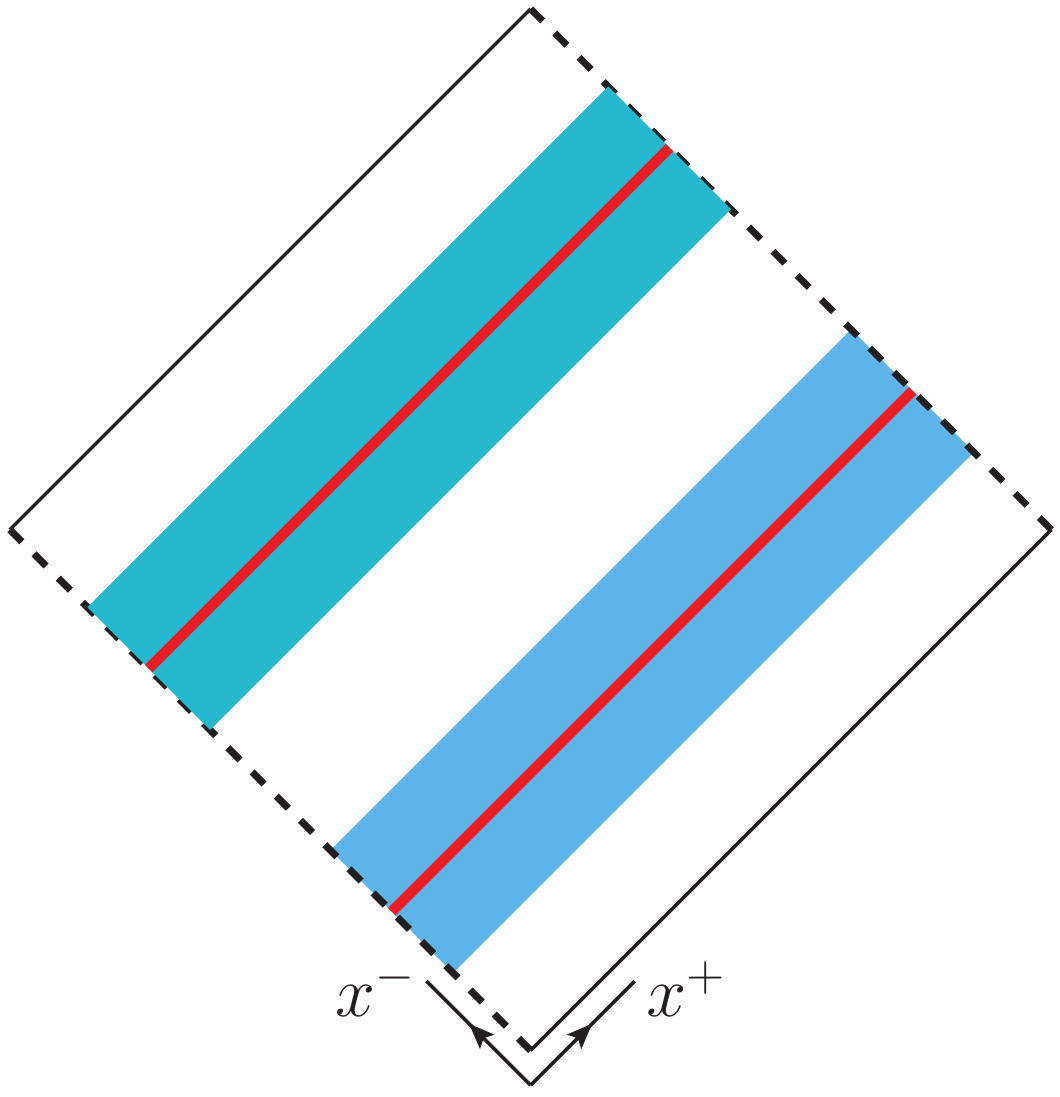}

    \caption{\label{fig:twopulse}
    Illustrations of symmetric oscillating fields such that $b_\infty$, defined in (\ref{eq:definition-b}), is zero. The associated potentials and horizons are also shown.}
\end{figure}

Consider the wave-like background as in the left hand panel of Fig.~\ref{fig:twopulse}, roughly two copies of our original field, back to back and of opposite sign. We choose signs such that the `first' part of the field (that at lower $x^\LCm$, as would be encountered first by an incoming particle) has $e\mathcal{E}>0$, as before. For a symmetric wave, $b(x)$ as defined in (\ref{eq:definition-b}) achieves its maximum in the middle of the wave, then decreases back to zero.

Arguing as before, one finds that all positrons traverse this field, but now with zero impulse -- there is no velocity memory. There is a future horizon in the first part of the pulse for incoming electrons, as before. Those that can traverse the first part of the pulse can also traverse the second, again with net zero impulse. As such, it follows that the `smoking gun' of pair production is now the presence of \emph{electrons} with $0< p_\LCp < b_\text{max}$ in the out-region, as these are classically forbidden. There is a past horizon for these electrons in the \emph{second} part of the field. If we calculate the pair production amplitude as before, then using either boundary between the field and the flat regions, the answer is zero.

To explore this result, we imagine that the two parts of the wave are drawn apart, becoming temporally separated in $x^\LCm$ as in the right hand panel of Fig.~\ref{fig:twopulse}. The spacetime is vacuum between the two patches of field.
Repeating our previous calculations for this field, there are nonzero contributions only from the boundaries with vacuum \emph{between} the two patches of field. These two contributions are exactly equal, up to a minus sign coming from the orientation of the two boundaries, and so the pair production amplitude vanishes. 

The arguments above seem to hold not just when we consider the particular case of a highly symmetric pulse, but whenever the field obeys $b_\infty=0$; contributions between different patches of the oscillating field cancel out. If, on the other hand, the field is oscillating but $b_\infty\not=0$, then there will typically be a non-zero contribution on the initial boundary, as for our initial example of $e\mathcal{E}>0$. The field may have a single or multiple horizons for different momenta depending on its profile. In summary, the result is that one requires $b_\infty\not=0$ in order for there to be pair creation. Equivalently, there is only observable particle production in the quantum theory if there is classical velocity memory. This is intriguing, but, since there are ambiguities when crossing horizons, it is fair to say that this point warrants further investigation.

\subsection{Radiation and the classical limit}
We will consider here how the classical radiation results of Sect.~\ref{sect:classical-R} arise from amplitudes in our system. There are two motivations for this. First, the interest in obtaining classical gravitational observables from amplitudes, as discussed in the introduction. Second, the fact that our system is only usually considered in the context of pair creation; to our knowledge other scattering amplitudes have not been studied in any detail.

The leading-order radiation reaction/self force effect in the system is the emission of radiation from a probe particle. In the quantum theory this is encoded in the three-point amplitude $\mathcal{A}_3(p\to q+ k^s)$ (on the background) for a particle of initial momentum $p$ to scatter to momentum $q$, emitting a photon of momentum $k$ and helicity~$s$. In terms of this amplitude, the total radiated power in the quantum theory is~\cite{Krivitsky:1991vt,Higuchi:2002qc,Ilderton:2013tb,Kosower:2018adc}
\be\label{eq:Kkvant}
        \langle K_\mu \rangle = \sum\limits_s \int\!\ud k_\text{o.s.}\!\int\!\ud q_\text{o.s.}\, k_\mu |\mathcal{A}_3(p\to q+ k^s)|^2 \;.
\ee
Returning, from here on, to the initially studied field $e\mathcal{E}>0$, we take our incoming particle to be an electron so that the horizon is in play.  In the perturbiner approach the amplitude (\ref{eq:Kkvant}) is computed by the three-point interaction term of (\ref{scalar-action}), evaluated on incoming and outgoing electron wavefunctions, and a free photon wavefunction: 
\be\label{eq:3}
    \mathcal{A}_3(p\to q+ k^s) = ie\int\!\ud^4 x\, e^\LCm_\text{out}(q;x)\, e^{ik\cdot x}\varepsilon(k)\cdot \overset{\leftrightarrow}{\partial}_x e^\LCm_\text{in}(p;x) \;,
\ee
where we have chosen the gauge $n\cdot\varepsilon=0$. We focus on the more interesting case of $p_\LCp <b(x^\LCm)$ so that there is a future horizon for the incoming electron. (The other case will follow trivially.) The amplitude contains two contributions, one from before the horizon, $x^\LCm < b^{-1}(p_\LCp)$, and one from beyond it, $x^\LCm > b^{-1}(p_\LCp)$. The former contribution is
\be\label{eq:A3}
\begin{split}
    &2e {\hat \delta}^2(q_{\LCperp}+ k_{\LCperp} -p_{\LCperp})
    {\hat \delta}(q_{\LCp}+k_{\LCp} -p_\LCp + b_\infty)
    \\
    &\int_{-\infty}^{b^{-1}(p_\LCp)}\!\ud x^\LCm
    \varepsilon\cdot\pi(x^\LCm)
    \bigg(\frac{p_\LCp (p_\LCp-b_\infty - k_\LCp)}{\pi_\LCp(x^\LCm) (\pi_\LCp(x^\LCm) - k_\LCp)}\bigg)^{1/2} \exp\bigg[i \int_0^{x^\LCm}\!\!\ud y\, \frac{k\cdot\pi(y)}{\pi_\LCp(y)-k_\LCp}\bigg]\;.
\end{split}    
\ee
The leading classical behaviour of this part of the 3-point amplitude is straightforward to extract. Photon momenta scale with $\hbar$~\cite{Krivitsky:1991vt,Higuchi:2002qc,Holstein:2004dn,Ilderton:2013tb,Kosower:2018adc}, so we may drop factors of $k$ \emph{relative} to factors of massive momenta. This leaves:
\be\label{eq:under}
\begin{split}
    & 2e\, {\hat \delta}^2(q_{\LCperp} -p_{\LCperp})
    {\hat \delta}(q_{\LCp} -p_\LCp + b_\infty)
    \\
    &\int_{-\infty}^{b^{-1}(p_\LCp)}\!\ud x^\LCm
    \varepsilon\cdot\pi(x^\LCm)
    \frac{\sqrt{p_\LCp (p_\LCp-b_\infty)}}{\pi_\LCp(x^\LCm)}\exp\bigg[i \int_0^{x^\LCm}\!\ud y\, \frac{k\cdot\pi(y)}{\pi_\LCp(y)}\bigg] \;,
\end{split}    
\ee
as the leading classical contribution. Before processing this further, we observe that the second contribution to (\ref{eq:3}), from beyond the horizon, will be largely similar, except that it will carry the Schwinger exponent resulting from continuing the incoming wavefunction. So, reinstating $\hbar$ explicitly, this contribution will be proportional to
\be
      \exp \bigg[-\frac{\pi m^2}{e\hbar\, \Epsilon(b^{-1}(p_\LCp))} \bigg]\;.
\ee
In the classical limit the $1/\hbar$ in the exponent kills this contribution non-perturbatively: there is nothing else in the amplitude that can compensate for this. Hence we can drop the contribution from beyond the horizon entirely. This means simply that classical radiation is sourced entirely from that part of the classical trajectory before the horizon, which makes sense since the horizon defines the asymptotic limit of the particle trajectory. The simplifications afforded by the classical limit, in going from (\ref{eq:A3}) to (\ref{eq:under}), are easy to interpret; dropping $k$ in the delta functions means that momentum conservation is that of Lorentz force (geodesic) motion. Hence we obtain the radiation emitted by a test particle on a geodesic -- this is the expected first self force/radiation reaction correction. The rest of the expression is proportional to the classical current\footnote{Phases associated to the lower limit of the integral in (\ref{eq:under}) correspond to a choice of asymptotic initial position, akin to an impact parameter. In our system the radiated power is independent of this parameter.}:
\be\label{eq:under2}
    \mathcal{A}_3 \to 2\, {\hat \delta}^2(q_{\LCperp} -p_{\LCperp})
    {\hat \delta}(q_{\LCp} -p_\LCp + b_\infty)\sqrt{p_\LCp (p_\LCp-b_\infty)} \varepsilon^\mu(k) j_\mu(k) \;.
\ee
Mod-squaring the amplitude, the product of delta functions is regulated by including a wavepacket for the initial state. Since we have conservation of three-momentum this factorises out. The remaining delta functions are eliminated by the $\ud q_\text{o.s.}$ integral. This sum over asymptotic states also removes the square root factors in (\ref{eq:under2}), and summing over photon helicities leaves
\be\label{eq:K2}
	\lim_{\hbar\to 0} \,\langle K_\mu \rangle = \int\!\ud k_\text{o.s.}\, k_\mu\, j^\dagger_\nu(k) j^\nu(k) \;,
\ee
where $k$ is now the wavenumber of the classical radiation field. This matches the classical expression\footnote{We note in passing that without the prefactor in our wavefunctions, i.e.~the next-to-leading order (NLO) WKB term, the three-point amplitude does not reduce to the correct classical current as in (\ref{eq:under2}), nor is the radiated power (\ref{eq:K}) recovered. This illustrates the general statement that it is not the \emph{geometric optics} approximation (leading order WKB), but rather the \emph{physical optics} approximation (the leading and NLO WKB terms) which constitutes the leading asymptotic behaviour of the approximated function~\cite[\S 10]{Bender}.} (\ref{eq:K}), and thus the explicit result (\ref{eq:K-explicit}). The remaining cases of electrons with $p_\LCp>b_\infty$ or any positron, where there is no horizon, goes through just as for (\ref{eq:A3})--(\ref{eq:under}).

\subsection{Spinor pair production}
We end our amplitudes discussion by extending the pair creation calculation to the case of spinor QED. To do so we need to solve the Dirac equation in the field (\ref{TheField}), for which there is a very simple method. Observe that, classically, (\ref{TheField}) simply boosts particles in the $z$--direction.  Following~\cite{Chiu:2017ycx}, we therefore work in a basis of free lightfront helicity spinors $u^\sigma_p$, as these have the property that the helicity is (i) equal to the $z$-component of spin in the rest frame, (ii) \emph{invariant} under the product of lightfront boosts $M_{\LCp\LCm}$ then $M_{\LCm\LCperp}$ which take us from rest to an arbitrary momentum and (iii) thus invariant under the boost imparted by (\ref{TheField}). (Explicit expressions for these spinors are given in Appendix~\ref{app:helicity}.) Since the scalar wavefunctions above are almost semiclassical-exact, we guess that the Dirac equation is solved simply by multiplying them by the relevant spinor ($u_p$, $v_p$, ${\bar u}_p$ or ${\bar v}_p$) with $p_\mu$ replaced, component-wise, by the appropriate classical momentum $\pi_\mu$. This guess is correct (and a direct solution reveals the same result), so for example the incoming electron is simply
\be\label{eq:spinor-wavefunction-in}
    \psi_\text{in}(p;x) = u^\sigma_{\pi(x)} e^\LCm_\text{in}(p;x) \;.
\ee
The perturbiner expression for the QED pair creation amplitude is then
\be\label{eq:todoQED}
\mathcal{A}_\text{QED} = \int \ud^2 x^\LCperp \ud x^\LCp \, e^{i q \cdot x} {\bar u}^\sigma_q \gamma^\LCm \, v^{\sigma'}_{\pi(x^\LCm)} e^\LCp_\text{out}(x) \bigg|_{x^\LCm = x^\LCm_i},
\ee
which, it can be checked, is in the form one expects from LSZ reduction on a surface of constant $x^\LCm$.  The spinor product in (\ref{eq:todoQED}) is, in our representation (see Appendix~\ref{app:helicity}),
\be
{\bar u}^\sigma_q \gamma^\LCm v^{\sigma'}_{\pi(x^\LCm)} = 2 \sqrt{q_+ (p_\LCp-b_\infty)} \,\big(1 - \delta^{\sigma \sigma'}  \big),
\end{equation}
which is non-zero only for pairs of opposite spin. The spinor amplitude differs only from the scalar amplitude (\ref{amplitudes}) by a spin factor and an irrelevant phase:

\begin{equation}
 		\mathcal{A}_\text{QED} = -i \big(1 - \delta^{\sigma \sigma'}\big) \mathcal{A}_{\text{SQED}},
\end{equation}
Squaring up, we sum over spins, which yields a factor of two relative to the scalar case, simply accounting for the number of spin states that can be created. The integrals over the pair momenta are the same as in scalar QED, and we obtain 
\begin{equation}
\label{eq:final2}
\begin{split}
    \mathbb{P}_\text{QED} &= 2\mathbb{P}_\text{SQED}\;,
\end{split}
\end{equation}
which again agrees with the literature results~\cite{Tomaras:2000ag,Tomaras:2001vs,Fried:2001ur,Ilderton:2014mla}. 

\section{Conclusions}\label{sect:concs}
%
We have discussed an electromagnetic `analogue' of a causal horizon. This is a toy model, but it exhibits several effects commonly associated with gravitational horizons, and provides a simpler arena in which to examine them. Classically, the horizon is of Rindler-type, realised through non-uniform acceleration. Quantum mechanically, there is pair production at the horizon due to the Schwinger effect, from which only one particle escapes to infinity. Unitarity is violated when crossing the horizon, and there is no well-defined $S$-matrix. Despite this, we have used the perturbiner approach to extract the pair creation probability and the power emitted by particles scattering on the field.

In future work it would be interesting to see if higher loop and higher multiplicity results could be found in the quantum theory, and how these exponentiate to recover the all-orders classical results such as (\ref{RR}) in Sect.~\ref{sect:RR}. The methods of ~\cite{Torgrimsson:2021wcj} seem well-suited to this task. Another topic to pursue would be the interaction of more complex initial states with the background, and the horizon. For example, beginning with a squeezed state (a coherent state of particle-antiparticle pairs), its interaction with the background would allow one to explore interference between pairs initially present and those created by the field. Such an interaction is intimately tied to the (in-) stability of the Bunch-Davies vacuum in de Sitter~\cite{Polyakov:2007mm,Anderson:2013ila,Anderson:2013zia}, and indeed an electromagnetic analogue of this interaction has already been discussed for constant fields in~\cite{Anderson:2017hts}. Studying the model here in the same context could be particularly interesting because quantum interference effects in the Schwinger process become very rich in non-constant fields~\cite{Akkermans:2011yn}.
    
It would also be extremely interesting to investigate the double copy of the results presented here.  There are many open questions regarding the extent to which double copy applies beyond perturbation theory and beyond flat space, and results from solvable models, such as that discussed here, may help to gain insight into colour kinematic duality, BCJ relations~\cite{Bern:2008qj,Bern:2010ue} and double copy beyond perturbation theory, or in curved geometries~\cite{Adamo:2017nia,Cheung:2021zvb,Sivaramakrishnan:2021srm,Zhou:2021gnu,Diwakar:2021juk,Armstrong-Williams:2022apo,Cheung:2022pdk}. It would be particularly interesting to establish explicitly how the double copy of \emph{non}-perturbative Schwinger pair production is related to  Unruh and Hawking radiation~\cite{Kim:2016dmm,Volovik:2022cqk}. {Such investigations could begin at the level of classical double copy, i.e.~ by looking at solutions to the Einstein and Maxwell equations. If the relevant spacetime metric fits into the Kerr-Schild class, then the method proposed in~\cite{Chawla:2023bsu} would allow one to identify the location of the horizon. The fact that our gauge field is sourced will likely become relevant here~\cite{Easson:2021asd}.} Further, we know our system is essentially two dimensional, and here some non-perturbative double-copy results are already known~\cite{Cheung:2022mix}. One can also ask what the single copy of horizon effects are in gauge theory; again, our system may offer a setting in which to explore this.

Work in these directions will be reported elsewhere. \\

\noindent\textit{We thank Tim Adamo, Suddho Brahma, Riccardo Gonzo, Tom Heinzl and Donal O'Connell for useful discussions and feedback on the manuscript. W.L.~is supported by an STFC studentship.}

\appendix

\section{Lightfront helicity spinors}\label{app:helicity}

In order to construct the QED wavefunctions in the text we use a basis of free lightfront helicity spinors (which solve the Dirac equation in the absence of external fields). As discussed in \cite{Chiu:2017ycx}, this basis encompasses a boost invariant notion of spin. The definition begins with the following choice of classical spin vector $s_p^\mu$ for a particle of momentum $p$:
\be
    s^\mu_p := \frac{1}{m}\bigg(p^\mu - \frac{m^2}{n\cdot p}n^\mu \bigg) \;,
\ee
where $n^\mu$ is as in the text. We have $s_p\cdot p=0$ and $s_p^2 = -1$. From the Pauli-Lubanski pesudovector $W_\mu$,
\be
    W_\mu = -\frac12 \epsilon_{\mu\nu\zeta\rho} P^\nu M^{\zeta\rho} \;,
\ee
with $P$ the momentum operator and $M^{\zeta\rho} = (i/4) [\gamma^{\zeta},\gamma^{\rho}]$ the Lorentz generator, we define $L_p = 2 s_p \cdot W /m$. Lightfront helicity spinors $u_{p,\sigma}$ with $\sigma=\pm 1$ are then defined by 
\be
    L_p u_{p}^{\sigma} = \frac{1}{m} \gamma_5 \slashed{s}_p \slashed{p} \, u_{p}^{\sigma} = \sigma\, u_{p}^{\sigma} \;. 
\ee
To present the explicit spinors we work in the Weyl basis of $\gamma$-matrices -- our convention are as in~\cite{schwartz2013}. With normalisation $\bar u u = 2m$ the basis of spinors is
\begin{equation}
\begin{split}
    u_p^{-1} =& \frac{1}{2^{1/4} \sqrt{p_\LCp}}
     \begin{bmatrix}
         0 \\
         m \\
         -p_1 + i p_2 \\
         \sqrt{2} p_\LCp
     \end{bmatrix}, \;
     \qquad
     u_p^{+1} = \frac{1}{2^{1/4} \sqrt{p_\LCp}} 
    \begin{bmatrix}
        \sqrt{2} p_\LCp \\
        p_1 + i p_2 \\
        m \\
        0
    \end{bmatrix},
     \\
     v_p^{-1} =& \frac{1}{2^{1/4} \sqrt{p_\LCp}}
     \begin{bmatrix}
         0 \\
         -m \\
         - p_1 + i p_2 \\
         \sqrt{2} p_\LCp
     \end{bmatrix},
     \;
     \qquad 
     v_{p}^{+1} = \frac{1}{2^{1/4} \sqrt{p_\LCp}}
     \begin{bmatrix}
         \sqrt{2} p_\LCp \\
         p_1 + i p_2 \\
         -m \\
         0
     \end{bmatrix}.
\end{split}
\end{equation}
See also~\cite{Ilderton:2020gno} for an application of these spinors in a different electromagnetic system.

\section{Zero modes}\label{app:zeromodes}
Consider again the incoming scalar electron (\ref{electron-in}) and define the function
\be\label{eq:ZM1}
    \varphi(x) := \Theta[p_\LCp -b(x^\LCm)] \, e^\LCm_\text{in}(p;x) \;.
\ee
for $\Theta$ the Heaviside step function. This function is zero beyond the horizon $x^\LCm_c = b^{-1}(p_\LCp)$. If we plug this into the KG equation we find, using the fact that $e^\LCm_\text{in}$ is a solution
\[
\begin{split}
    \big( \mathcal{D}^2 +m^2 \big) \varphi(x) &= 2\big(\partial_\LCm \Theta[p_\LCp-b(x^\LCm)] \big)\partial_\LCp e^\LCm_\text{in}(p;x)  \\
    &= -2i e\mathcal{E}(x^\LCm) \delta(p_\LCp - b(x^\LCm)) \sqrt{p_\LCp-b(x^\LCm)} \exp\cdots \;,
\end{split}
\]
in which "$\exp$" stands for all exponential terms in (\ref{electron-in}). This is zero when multiplied by test functions, albeit weakly because the distribution is $\sqrt{p}\, \delta(p)$ rather than, say, $p\, \delta(p)$ (but see below), and so (\ref{eq:ZM1}) is a solution of the KG equation.

This arguments extend directly to all wavefunctions with a horizon.
This means that we can turn off, or turn \emph{on}, any of our functions at their horizon, and still have a solution of the KG equation. 
Given e.g.~our incoming electron, there is seemingly nothing in our characteristic, or lightfront, setup which determines exactly what should or should not be added. This makes it clear that evolution across the horizon is not unitary.  From the perspective of lightfront field theory these solutions manifest the infamous `zero modes', which are a long-standing issue~\cite{Mccartor:1988bc,Heinzl:1991vd,Ji:1995ft,Heinzl:2003jy}. See~\cite{Mannheim:2019lss,Mannheim:2020rod} for recent discussions.

The spinor story is only a little different. To see how the constraining step function should be imposed it is best to use a 2-spinor formalism. To that end we introduce the orthogonal projectors $\Lambda_\LCpm := \frac12 \gamma^\LCmp \gamma^\LCpm$ and define $\psi_\LCpm = \Lambda_\LCpm \psi$. The full field may then be written $\psi = \psi_\LCp + \psi_\LCm$. For simplicity of presentation we strip the factors of $\exp -i p_\LCperp x^\LCperp$ from all wavefunctions (since these are not affected by the background). The Dirac equation becomes, when multiplied by the projectors, equivalent to
\be\label{eq:2-spinor-Dirac}
\begin{split}
    \partial_\LCp \mathcal{D}_\LCm \psi_\LCm &= - \frac{m_\LCperp^2}{2} \psi_\LCm \;,
    \qquad
    \psi_\LCp = \frac{p_\LCperp \gamma^\LCperp +m}{m_\LCperp^2} i \mathcal{D}_\LCm \gamma^\LCm \psi_\LCm  \;,
\end{split}
\ee
where $m_\LCperp^2 := p_\LCperp^2 +m^2$. The second of (\ref{eq:2-spinor-Dirac}) is a constraint, determining $\psi_\LCp$ if $\psi_\LCm$ is known. We now proceed as for the scalar case. 
Using incoming electrons to illustrate, we project out $\psi_\LCm$ from our existing solution (\ref{eq:spinor-wavefunction-in}) and multiply by the same step function. This defines a new function $\chi_\LCm$, and $\chi_\LCp$ is then fixed by the second equation of (\ref{eq:2-spinor-Dirac}). Acting with the Dirac equation, we now find

\be
    \big(i\slashed{\mathcal{D}}+m\big) \chi \sim (p_\LCp - b(x\LCm)) \delta(p_\LCp - b(x\LCm)) \exp\cdots
\ee
and so we now have a proper distributional zero, $p\, \delta(p)$, and these are solutions of the Dirac equation. Again, it follows that one can add any multiple of these functions to a given solution, at the horizon, and unitarity is violated. 

\section{Alternative approaches}\label{app:TTW}
Starting from our background field (\ref{TheField}), consider truncating is so that it covers only a semi-infinite range $x^\LCp>L$, so
\be
    A_\LCm = -x^\LCp \mathcal{E}(x^\LCm)\to -\Theta(x^\LCp-L)(x^\LCp-L) \mathcal{E}(x^\LCm) \;.
\ee
This is the system analysed in~\cite{Tomaras:2000ag,Tomaras:2001vs}. Unlike in our field, free data can here be specificed on \emph{both} characteristics of the theory, $x^\LCpm = $constant.  Using canonical methods, a massive spinor field was quantised on both characteristics, $x^\LCm=x^\LCm_i$ and $x^\LCp=L$ (equivalent to any two such null surfaces before the background turns on), and evolved in a two-time formalism into the region $x^\LCm>x^\LCm_i$, $x^\LCp>L$.

The singularities in the corresponding spinor wavefunctions remain, but the extra boundary data has two important consequences. First, it dictates in which direction to go around singularities in the complex plane. Second, it generates additional terms in the wavefunctions, which switch on at the horizons, and which maintain unitarity. Essentially, a prescription is giving for dealing with the zero mode solutions discussed in Appendix~\ref{app:zeromodes}.

Ambiguities are thus removed by modifying the system such that \emph{Cauchy} data, rather than characteristic data, can be specified. However, this approach cannot be applied directly to our field (\ref{TheField}) when the `full' horizon is in place, as then there is no second characteristic at constant $x^\LCp$ on which to specify free data. One can take the $L\to-\infty$ limit of the results in~\cite{Tomaras:2000ag,Tomaras:2001vs}, but the explicit form of the spinor wavefunctions in that limit shows that information from the second surface remains; one is still quantising on two characteristics, and our solutions (\ref{eq:spinor-wavefunction-in}) are not recovered. This is not surprising; if information is specified on two surfaces, it should not be lost just by moving one surface to asymptotic distance. If contributions from the second surface are dropped, information is lost, and this results in the loss of unitarity at the horizon.

In our system, we are left with the question of whether it is possible to calculate without using a two-time formalism. We note in this context that the pair production probability (or strictly the vacuum decay probability) has been rederived using path integral~\cite{Fried:2001ur} and worldline~\cite{Ilderton:2014mla} methods, without having to specify data on two characteristics (at least not explicitly). This prompts the question of whether there are ways to set up scattering calculations in order to obtain the correct physical results.   
An alternative perspective is to ask whether it is possible to replace Cauchy data by ``characteristic data plus some prescription'', which still allows one to quantise on only one characteristic. This idea was explored in~\cite{Heinzl:1993px}, in which it was shown that quantisation on a single characteristic (e.g.~$x^\LCm$) can be consistent if one in addition imposes periodic boundary conditions in the other characteristic direction ($x^\LCp$). In this paper, we identified another prescription, based on the perturbiner approach, and confirmed that we could use it to compute the correct pair production probability.

\bibliographystyle{JHEP}

\providecommand{\href}[2]{#2}\begingroup\raggedright\endgroup

\end{document}